\documentclass[12pt]{iopart}
\usepackage{calrsfs}
\usepackage{varwidth,xcolor}
\usepackage{graphicx}
\usepackage{subfig}
\usepackage{float}
\DeclareGraphicsExtensions{.pdf,.jpeg,.png,.eps,.jpg}
\begin{document}

\title[Entropy production and rectification efficiency]{Entropy production and rectification efficiency in colloids transport along a pulsating channel}

\author{M. Florencia Carusela}
\address{Instituto de Ciencias, Universidad Nacional de General Sarmiento, J.M.Gutierrez 1150, CP 1163, Los Polvorines, Buenos Aires, Argentina; \\ CONICET, Argentina}
\ead{flor@ungs.edu.ar}

\author{J. Miguel Rubi}
\address{Departament de F\'isica de la Materia Condensada, Universitat de Barcelona, C/Mart\'i Franqu\`es 1, 08028, Barcelona, Spain}

\begin{quotation}

We study the current rectification of particles moving in a pulsating channel under the influence of an applied force. We have shown the existence of different rectification scenarios in which entropic and energetic effects compete. The effect can be quantified by means of a rectification coefficient that is analyzed in terms of the force, the frequency and the diffusion coefficient. The energetic cost  of the motion of the particles expressed in terms of the entropy production depends on the importance of the entropic contribution to the total force. 
Rectification is more important at low values of the applied force when entropic effects become dominant. In this regime, the entropy production is not invariant under reversal of the applied force. 
The phenomenon observed could be used to optimize transport in microfluidic devices or in biological channels.

% \cite{nuestro}.

\end{quotation}

\date{}

\section{Introduction}

Particle current rectification is an important mechanism that control transport at small scales. It is based on the breaking of the intrinsic randomness of Brownian fluctuations thus facilitating the motion of the particles in a preferred direction \cite{Reimann,hangi}. The mechanism takes place when particles moves in an asymmetric (ratchet) potential in whose case detailed balance is not fulfilled and a net current comes up \cite{bartu, machura}. 

Current rectification has also been predicted in transport through confined structures having irregular boundaries \cite{jacobs, zwan,rubi7,kalinay,rubi8,rubi9,rubi4b,marquet,rubi2,dad,dad3}. In this situation, the entropy (a function of the number of states accessible to the particles) is not a constant along the transport direction and entropic forces related to the entropy gradient act on the particles. Transport in the presence of entropic barriers, or entropic transport, plays an important role at the mesoscale and has been subject of many studies \cite{ion,Lairez,rubi6,holu,dorf}.

A new entropic rectification mechanism has recently been proposed for the case in which the channel through which particle move undergoes periodic deformations. Periodic changes of the channel structure result in time-dependent entropic barriers that gives rise to peculiar transport properties. It has been found that channel pulsations may induce current reversal and resonant effects \cite{nuestro,entropic,entropicw}.

In this article, we study the efficiency of the entropic rectification process in a pulsating channel by quantifying the phenomenon through a rectification coefficient that is analyzed in terms of the force applied, the height of the barrier and the diffusion coefficient. The coefficient characterizes the different rectification regimes. Rectification is also studied under the prism of the energy dissipation in the process, or equivalently of the entropy production. We will show that at high values of the applied force the entropy production is practically invariant under inversion of the force. At small values of the force, when entropic effects are dominant, the symmetry is broken.

The article is organized as follows. In Section 2, we briefly review the entropic transport model for a pulsating channel. Section 3 is devoted to the analysis of the rectification coefficient. In Section 4, we analyze the energetic cost of the rectification mechanism by computing the entropy production from the knowledge of the particle current and the force. Finally, in Section 5 we present our main conclusions.

%modelled by means of entropic barriers [], [],  may give rise to a geometric or entropic bias of the Brownian fluctuations.  Entropy variations arising from confinement can thus break down the intrinsic randomness of the fluctuations and can make that moves in the direction opposed to the entropy gradient, from more to less confined regions. This fact gives rise to an entropic rectification mechanism [] that can explain why ions can proceed through protein channels. Channels can thus be designed to optimize particle transport [].
%A different type of rectification is observed when the shape of the region in which particles move varies in time. This case can be modelled by a time-dependent entropic potential When the entropic barriers depend not only on position but on time new rectification scenarios come up. The corresponding rectification mechanism has been reported in pulsating channels where current inversion and resonances were observed []. ....
%In this article, we contribute to the study of entropic rectification in a pulsating channel by quantifying the phenomenon by means of a rectification index which is analyzed in terms of the force applied, the height of the barrier and the diffusion coefficient. The index characterizes the different rectification regimes.

\section{Entropic transport model for pulsating channels}

\begin{figure}[h]
\centering
\includegraphics[width=0.5\textwidth]{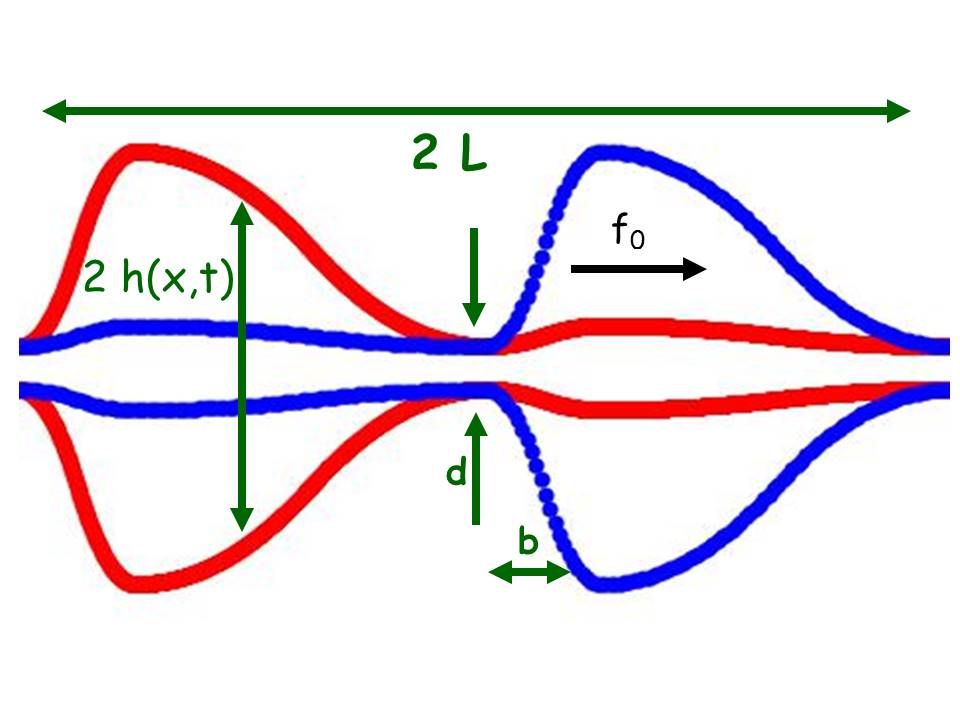}
\caption{Snapshot of a unit of the channel oscillating out of phase with period $T$ for two different times. The solid (black) line corresponds to $t=0$ and the dotted (blue) line corresponds to $t=T/2$.}
\label{fig:Figura1}
\end{figure}

We study the confined diffusion of $N$ non-interacting Brownian particles through a two-dimensional periodic channel whose walls may vary periodically in time. They consist of contiguous units of length $2 L$ formed by two subunits of length $L$, as is shown in Fig.\ref{fig:Figura1}

The shape of the boundaries of the channel is periodically modulated in time with the height given by:

\begin{equation}
h(x,t)= \left\{ \begin{array}{lcc}
             a_1(t) x^2 + \frac{d}{2} &   ;  &  0 \leq x \leq \frac{b}{2} \\
             \\ -a_1(t) (x-b)^2 + s(t)  &   ;  &  \frac{b}{2}< x  \leq  b\\
             \\ -a_2(t) (x-b)^2 + s(t)  &   ;  &  b < x  \leq \frac{L+b}{2}\\
						 \\ a_2(t) (x-L)^2 +   \frac{d}{2} &   ;  &  \frac{L+b}{2} < x \leq  L  
             \end{array}
   \right.
\end{equation}
Here $b$ indicates the location of the point of maximum width and $d$ is the width of the bottleneck. The time-dependent coefficients are $a_1(t)=\frac{2[s(t)-d/2]}{b ^2}$,  $a_2(t)=\frac{2[s(t)-d/2]}{(L-b)^2}$ and $s(t)=s_0+s_1 sin(\omega t+\Phi) $. The values of the parameters are set to guarantee the asymmetry of the subunits.  The phase difference between adjacent subunits in one unit cell is given by:

\begin{equation}
\Phi= \left\{ \begin{array}{lcc}
             0 &   ;  &   x \in [0,L]\\
						 \\ \pi   &   ;   &  x \in  (L ,2 L]
             \end{array}
   \right.
\end{equation}
A phase lag $\Phi=\pi$ means that consecutive subunits can shrink and enlarge alternatively such that the total volume does not change much, a realistic situation that may be observed in transport of particles through channels.

We will analyze the transport properties by means of the Fick-Jacobs equation, that governs the dynamics of the probability distribution of the ensemble of non-interacting Brownian particles

\begin{equation}
\frac{\partial P(x,t)}{\partial t}= \frac{\partial }{\partial x} \left[ D(x,t) \frac{\partial P(x,t)}{\partial x} -\frac{D(x,t)}{k_B T} F_{eff}(x,t) P(x,t) \right]
\label{FJ}
\end{equation}
Here $D(x,t)$ is an effective diffusion coefficient, that in our two-dimensional case is given by %\cite{rubi3}

\begin{equation}
D(x,t)= D_{0}{ (1+h'(x,t)^2)^{\left(-1/3\right)} }
\label{dif}
\end{equation}
where $D_0$ corresponds to the diffusion coefficient of the particles when they move in an unbounded medium and $\mathcal{F}_{eff}(x,t)$ is an effective force acting along the x-direction which is related to the energetic and entropic barrier contributions to the free energy $\mathcal{A}(x,t)$: 

\begin{equation}
F_{eff}(x,t)= -\frac{\partial A(x,t)}{\partial x}= F_0 + k_B T \frac{h'(x,t)}{h(x,t)}
\label{feff}
\end{equation}
with $A(x,t) \doteq E - T S = -F_0 x -k_B T \ln h(x,t)$.

From Eq.(\ref{FJ}), we can identify the instantaneous particle current
\begin{equation}
J(x,t)= - \left[D(x,t) \frac{\partial P(x,t)}{\partial x} -\frac{D(x,t)}{k_B T} F_{eff}(x,t) P(x,t) \right]
\label{J}
\end{equation}
The Fick-Jacobs approximation assumes that the probability density reaches equilibrium in the transverse direction much faster than in the longitudinal one. This requirement is fulfilled if $\left| h'(x,t) \right| << 1$ for all times and positions, that is, when the cross section of the tube varies smoothly.

For the sake of simplicity, we use dimensionless quantities.
We scale lengths with the unit length $L_o=2L$, times with the diffusion time $\tau_{dif}=L_o^2 \gamma/ (k_B T)$ with $\gamma$ the Stokes' friction of a spherical particle of reference radius $r$, energies with $k_B T$, forces with $k_B T /L_o$  and currents with $L_o/\tau_{dif}$. 
A typical diffusion constant in colloids in aqueous solution is $D \approx 10^{-12}m^2/s$. Therefore a typical Brownian time scale or average time for a particle to diffuse a distance equal to its diameter is of the order of $1 - 100$s, for particles of sizes from $1$ to $10\mu m$ and velocities in the range $10^{-1} - 1 (\mu m)/s$. %\cite{Dalle,Lopez}.
Besides, the validity of the Fick-Jacobs approach requires that the dimensionless frequency $\omega$ has to be smaller than one, this implies that modulations must be smaller than $20\pi$ rad/sec. These values are of the same order as the ones considered in recent experiments on transport of molecules in confined media subjected to entropic barriers and to a driving force \cite{Lairez}.

We can express the effective force in terms of dimensionless variables as:
\begin{equation}
\mathcal{F}_{eff}(x,t)= f_0\left(1 + \frac{1}{f_{0}} \frac{h'(x,t)}{h(x,t)}\right)
\label{fead}
\end{equation}
where the second term indicates the ratio between entropic and energy forces.
This equation shows that the effective force can equivalently be controlled by variations of the force, temperature and period of the cell. The effect of a small force would then be tantamount to the effect of the temperature.   

In addition to the conditions required for the validity of the Fick-Jacobs equation: smooth channel and slow oscillations, we assume that the viscosity of the fluid is high enough to justify the use of a diffusion equation strong viscous dynamics and that the suspension is dilute. Due to the latter assumptions, we can neglect hydrodynamic particle-wall and particle-particle interaction effects in the motion of the particles.

\section{Rectification efficiency}

From the dimensionless Fick-Jacobs equation, we obtain numerically the probability density $P(x,t)$ with periodic boundary conditions at $x=0, 1$. We introduce the mean particle current as

\begin{equation}
j(t) = \int^{1}_{0} J(x,t)  dx
\label{Jt}
\end{equation}
with the probability current $J(x,t)$ given by the dimensionless form of Eq.(\ref{J}). The average current $J$ is given by the time average of the particle current $j(t)$ in a time period $\tau=2\pi / \omega$ 

\begin{equation}
J = \frac{1}{\tau}\int^{\tau+t_0}_{t_0} j(t) dt
\label{Jt2}
\end{equation}

The particle current rectification strength achieved in the motion of colloids along the channel can be quantified through the rectification coefficient:

\begin{equation}
R=\frac{J(f_{0})-J(-f_{0})}{J(f_{0})+J(-f_{0})}
\end{equation}

\begin{figure}[H]
\minipage{0.5\textwidth}
\includegraphics[width=\linewidth]{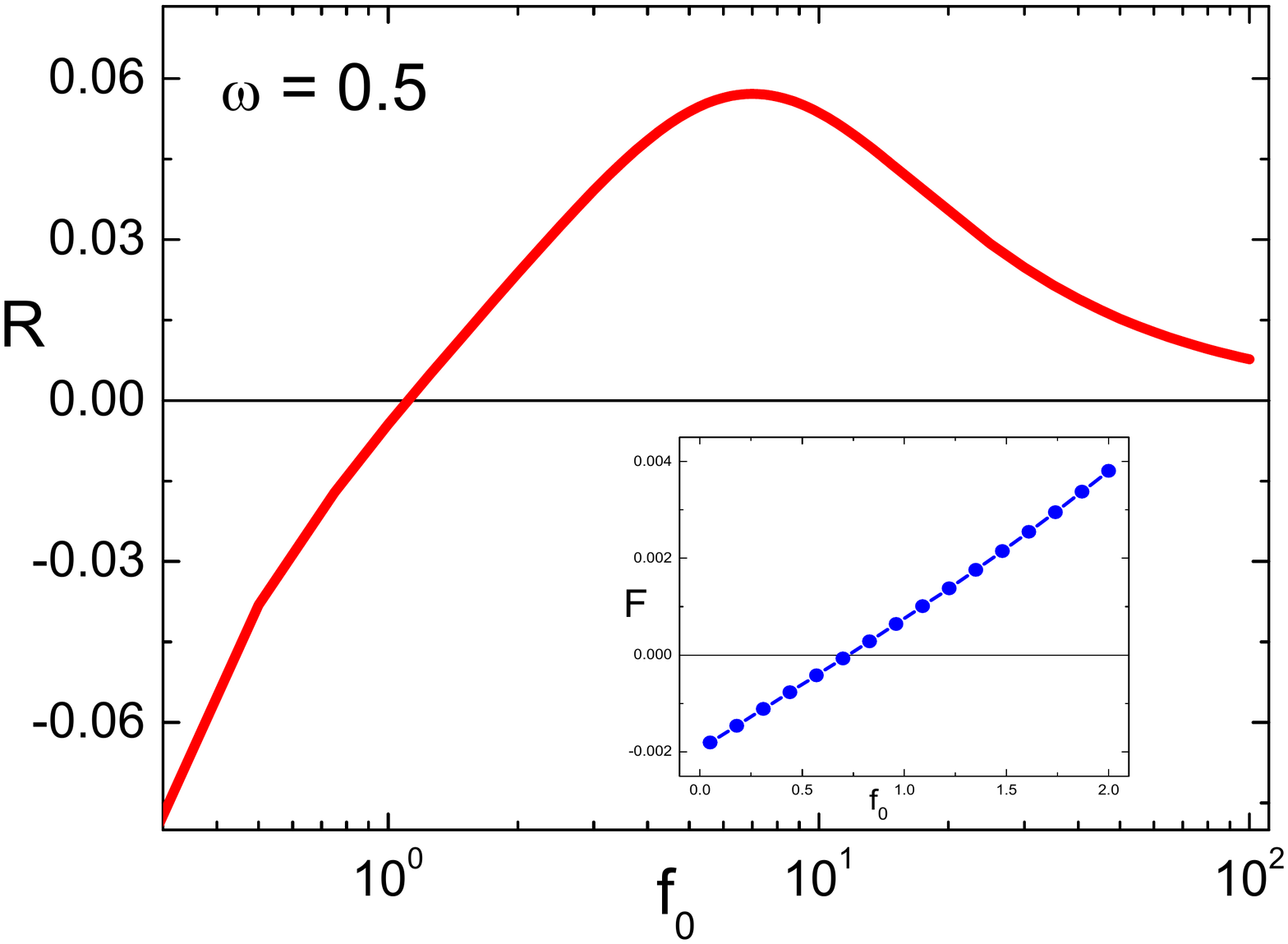}
\endminipage
\minipage{0.5\textwidth}
\includegraphics[width=\linewidth]{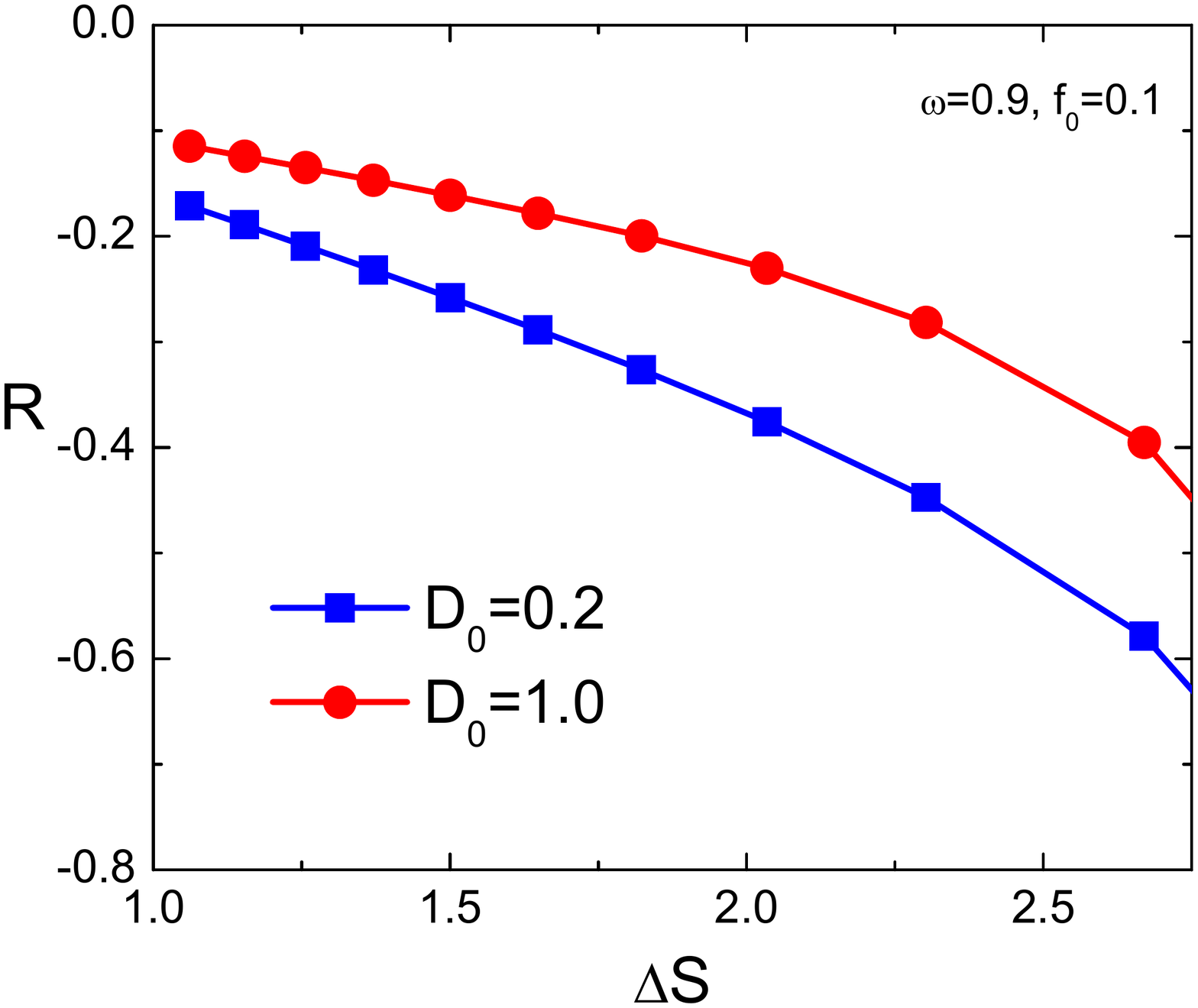}
\endminipage
\caption{Rectification coefficient as a function of the applied force and the entropic barrier. Left panel: R versus $f_{0}$ for $\omega=0.5$ and $\Delta_S=2.15$. Inset: Average mean force F versus  versus $f_{0}$. Right panel: R versus $\Delta S$ for low and high diffusion ($D_{0}=0.2, 1$). $\omega=0.9$ and $f_{0}=0.1$  Other parameters are: $s_0 = 0.45$, $s_1 = 0.2$, $b = 0.25$, $L=1$. }
\label{fig:Rw}
\end{figure}

\begin{figure}[H]
\centering
\includegraphics[width=0.7\textwidth]{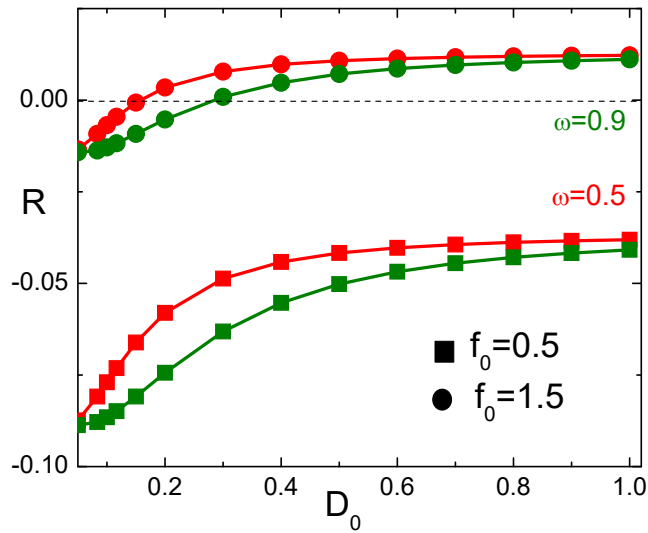}
\caption{Rectification coefficient R versus $D_0$ for $\omega=0.5$ (red) and $\omega=0.9$ (green). $f_{0}=1.5$ (circle), $f_{0}=0.5$ (square) and   $\Delta_S=2.15$ and other parameters as in Fig. \ref{fig:Rw}. }
\label{fig:RvsD}
\end{figure}

In the left panel of Fig.\ref{fig:Rw}, we observe a rectification of the particle current as a function of $f_{0}$. We find that R changes from negative to positive values as long $|f_0|$ increases. The inversion of R occurs for a critical value $f_{0}=f_{R}$, that in our case is $f_{R}\approx 1$. This phenomenon relies on a change over between a regime dominated by the entropic barrier to one dominated by the energy one. 
In the inset of the left panel of Fig.\ref{fig:Rw}, we plot the time average force F as a function of $f_{0}$ (positive), that is defined as:

\begin{equation}
F = \frac{1}{\tau}\int^{\tau+t_0}_{t_0} f(t) dt
\label{Jt3}
\end{equation}
with $f(t)$ the ensamble average of $\mathcal{F}_{eff}(x,t)$ at a given time.

For low external forces, F is negative due to the geometry of the channel. However as $f_0$ increases, F becomes positive.  The inversion of R can be understood from the behavior of F. We observe that the crossing of the horizontal axis occurs around the same value of $f_0$, for both R and F. For low but positive values of $f_0$, F is negative so the entropic forces dominates the transport.
For high and positive $f_0$, F becomes positive then the energy barrier dominates.  In other words, when the external force is very strong the dynamical response due to entropic forces is washed out. As the asymmetric contributions given by the geometry are negligible, the particle current takes the same values for a positive and negative $f_{0}$. This crossover relies on the relative importance between both contribution as it is expressed by the ratio in the second term of Eq.(\ref{fead}).

To show the competition of the different relevant time scales, we plot in Fig.\ref{fig:RvsD} the dependence of R on $D_{0}$ and $\omega$. 
In the energy dominated regime and low diffusion, R is more negative as long the pulsation is faster. In this situation, the time scale corresponding to the diffusion process is large and the faster time scale of the  entropic ratcheting dominates the dynamics. For larger $D_{0}$, diffusive time scale dominates and R saturates to a value independent of $\omega$. 

The situation is similar for low diffusion in the entropic regime, however R achieves a higher absolute value. That is, the entropy barriers enhance the role of the asymmetric confinement on the motion of colloids, effect that is washed out in the energy regime (larger $f_{0}$). For large diffusion, R tends to be independent of $\omega$ but always negative. 

In Fig.\ref{fig:RvsD} we observe a reversal of R when it is represented as a function of $D_0$.  As long $f_0$ increases, R becomes less negative and inversion takes place. The value of $D_0$ for which inversion occurs depends on $\omega$. As long the diffusion time is smaller, the channel has to oscillates faster to produce the same inversion phenomena. This fact suggests the presence of a scaling regime in which the reversal phenomena scales with $D_{0}/(L^{2}\omega)$, for a constant $f_0$.

The role played by the asymmetry of the channel is related to the entropic barriers. In the right panel of Fig.\ref{fig:Rw}, we plot R vs $\Delta S$, defined as the highest value achieved by the entropic barrier during its oscillation:  $\Delta S \equiv \ln(2(s_{0}+s_{1})/d)$.
We observe that in the entropic regime, rectification efficiency can be improved for larger $\Delta S$ and lower values of $D_{0}$.

\section{Entropy Production}
The different rectification scenarios found can also be analyzed in terms of the energetic cost of the particle transport which can quantified by means of the entropy production rate $\sigma(x,t)$. The dimensionless quantity (in units of $\tau_{dif}/k_{B}$) is the product of the flux $J(x,t)$ and the effective driving force $\mathcal{F}_{eff}(x,t)$:
\begin{equation}
\sigma(x,t) = -J(x,t)\mathcal{F}_{eff}(x,t)
\end{equation}

In the linear regime where the flux is proportional to the force the entropy production is proportional to the force squared and thus depends on the applied force and the entropic force or equivalently on the shape of the channel. The scaling behavior found for the effective force would then give rise to a scaling of the entropy production. The entropy production averaged over a period of time and space 
\begin{equation}
\sigma_{P} = \frac{1}{\tau} \int_{t_{0}}^{\tau +{t_{0}}} dt \int_{0}^{1} dx \hspace{0.2cm}\sigma(x,t)
\end{equation}
characterizes the dissipation inherent to particle transport.

\begin{figure}[h]
\centering
\includegraphics[width=0.7\textwidth]{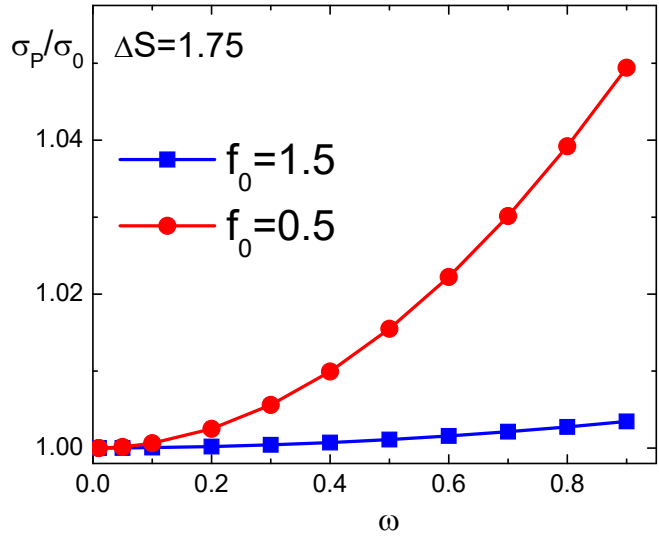}
\caption{Reduced entropy production: $\sigma_{P}(\omega)/\sigma_{0}(\omega \rightarrow 0)$ versus $\omega$, for $f_{0}=1.5$ (green) and $f_{0}=0.5$ (red). $\Delta_S=1.75$ and other parameters as in Fig. \ref{fig:Rw}. }
\label{fig:Swfo}
\end{figure}

\begin{figure}[H]
\centering
\includegraphics[width=0.7\textwidth]{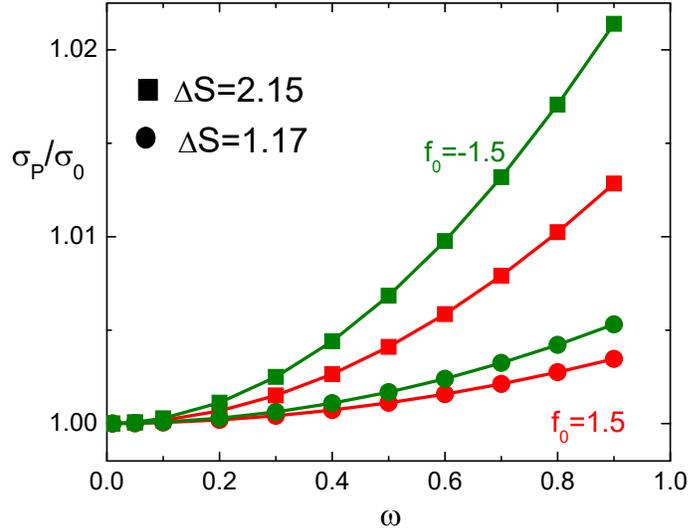}
\caption{$\sigma_{P}(\omega)/\sigma_{0}(\omega \rightarrow 0)$ versus $\omega$ for  $\Delta_S=2.15$ (squares) and  $\Delta_S=1.17$ (circles).  $f_{0}=1.5$ (red) and  $f_{0}=-1.5$ (green). Parameters as in Fig \ref{fig:Rw}. }
\label{fig:SwdS}
\end{figure}

\begin{figure}[H]
\centering
\includegraphics[width=0.7\textwidth]{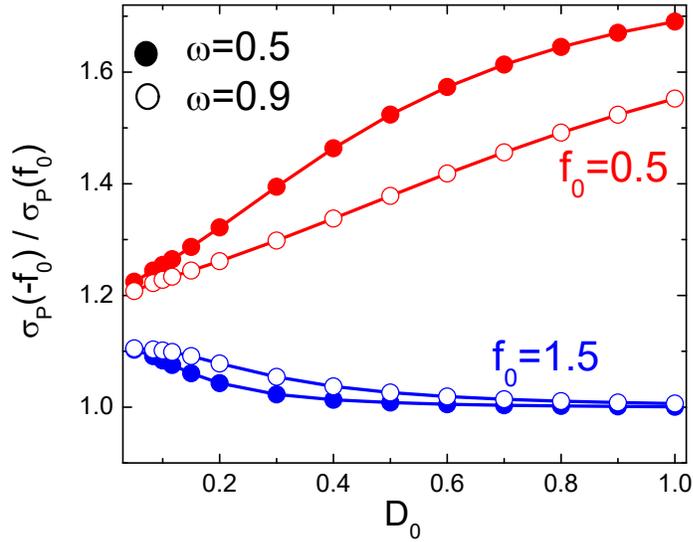}
\caption{$\sigma_{P}(-f_0)/\sigma_{p}(f_0)$ versus $D_0$ for $\omega=0.5$ (full circles) and $\omega=0.9$ (empty circles). $f_{0}=0.5$ (red), $f_{0}=1.5$ (blue) and $\Delta_S=2.15$ in all cases.}
\label{fig:difusion}
\end{figure}

In Fig.\ref{fig:Swfo}, we plot the entropy production normalized to the value obtained when $\omega \rightarrow 0$. Both quantities are positive and consequently the ratio is also positive for all frequencies.  For a given $\omega$, larger ratios are achieved in the entropic regime because the motion of the colloids is mainly ruled by the time-modulated entropic forces. In the entropic regime, the entropy production takes a value slightly larger than the static one, thus the dissipation increases.

In Fig.\ref{fig:SwdS}, we observe that a reversal of $f_0$ yields different values of the ratio $\sigma_{P}/\sigma_0$ that depends on the specific shape of the channel given by $\Delta S$. 

It is interesting to analyze how a reversal of the direction of $f_0$ affects the entropy production when the diffusion coefficient changes. In Fig.\ref{fig:difusion}, the ratio $\sigma_{P}(f_{0})/\sigma_{P}(-f_{0})$ is pictured as a function of $D_0$, for two different frequencies.  This figure clearly shows the presence of the two quite distinct regimes dominated by entropic and energy forces.

When the effect of the external force dominates, the left/right asymmetry vanishes, consequently the entropy production is the same in both directions. This effect is enhanced when the colloids diffuse faster (larger $D_0$) and tends to be independent of $\omega$. As long $f_0$ becomes larger, the system can be thought as colloids moving in a symmetric channel.

However, in the entropic regime the time scales related to frequencies and diffusion evenly play a relevant role. 
A diffusion time smaller than the characteristic time of the pulsation implies that the particle takes many time periods to travel a distance equal to one unit of the channel. Therefore, the effect of the time-dependent constraints imposed by the confinement is enhanced and the entropy production is larger, with different values for the forward and backward directions.

The behavior described suggests an interesting mechanism to induce transport of particles through microchannels. When external forces are small, the entropic ratcheting induced by pulsations becomes the leading mechanism in the transport process. It can be used to improve the rectification efficiency and to control the entropy production, reducing the dissipation.

\section{Conclusions}

In this article, we have analyzed the efficiency of the particle current rectification process observed in channels whose shape is modulated periodically. We have proposed a rectification coefficient that measures the difference between the particle current induced by a given force acting along both directions of the channel. We have identified two rectification regimes. At low values of the forces, when entropic effects become important, rectification may be relevant. On the contrary, at high forces the effect of the entropic barrier fades away and rectification considerably diminishes.   

We have examined the rectification regimes in terms of the entropy production which has been computed as a function of the applied force, the oscillation frequency the strength of the entropic barrier and the diffusion coefficient. The analysis of this quantity shows the existence of situations of minimum dissipation that can be selected upon varying these quantities. At high values of the force, the entropy production is invariant under inversion of the applied force. This invariance is broken at low forces when entropic effects become important. 

The results obtained indicate how channels should be designed for a controllable current and dissipation and why existing structures in nature such as protein channels and pumps undergo their main functions on the basis of their particular forms. 

\section{Acknowledgments}

This work is supported by CONICET-Argentina under Grant No. 14420140100013CO and MICINN of the Spanish Government under Grant No. FIS2015-67837-P.

%\bibliography{Biblio}

\end{document}